# Semantic Annotation and Querying Framework based on Semi-structured Ayurvedic Text

**Hrishikesh Terdalkar** **Arnab Bhattacharya**
hrishirt@cse.iitk.ac.in arnabb@cse.iitk.ac.in
Department of Computer Science and Engineering,
Indian Institute of Technology Kanpur,
India

**Madhulika Dubey** **Ramamurthy S** **Bhavna Naneria Singh**
drmadhulikadubey@gmail.com srmsvn@gmail.com bhavnasingh79@gmail.com
India

## Abstract

Knowledge bases (KB) are an important resource in a number of natural language processing (NLP) and information retrieval (IR) tasks, such as semantic search, automated question-answering etc. They are also useful for researchers trying to gain information from a text. Unfortunately, however, the state-of-the-art in Sanskrit NLP does not yet allow automated construction of knowledge bases due to unavailability or lack of sufficient accuracy of tools and methods. Thus, in this work, we describe our efforts on manual annotation of Sanskrit text for the purpose of knowledge graph (KG) creation. We choose the chapter Dhānyavarga from Bhāvaprakāśanighaṇṭu of the Ayurvedic text Bhāvaprakāśa for annotation. The constructed knowledge graph contains 410 entities and 764 relationships. Since Bhāvaprakāśanighaṇṭu is a technical glossary text that describes various properties of different substances, we develop an elaborate ontology to capture the semantics of the entity and relationship types present in the text. To query the knowledge graph, we design 31 query templates that cover most of the common question patterns. For both manual annotation and querying, we customize the Sangrahaka framework previously developed by us. The entire system including the dataset is available from `https://sanskrit.iitk.ac.in/ayurveda/`. We hope that the knowledge graph that we have created through manual annotation and subsequent curation will help in development and testing of NLP tools in future as well as studying of the Bhāvaprakāśanighaṇṭu text.

## 1 Introduction

Sanskrit (संस्कृत, IAST: Saṃskṛta) is one of the most prolific languages in the entire world, and text in Sanskrit far outnumber other classical languages. Consequently, with the advancement of natural language processing with the aid of computers, there has been a surge in the field of computational linguistics for Sanskrit over the last couple of decades. This has resulted in development of various tools such as the *Samsaadhanii* by Kulkarni (2016), *The Sanskrit Heritage Platform (SHP)* by Goyal et al. (2012), *Sanskrit Sandhi and Compound Splitter (SSCS)* by Hellwig and Nehrdich (2018), *Sanskrit WordNet (SWN)* by Kulkarni et al. (2010), etc. for linguistic tasks such as word segmentation, lemmatization, morphological generation, dependency parsing, etc. Despite this, many fundamental processing tasks such as anaphora resolution and named entity recognition that are needed for higher-order tasks such as discourse processing, are either not available or have a long way to go. Combined with the fact that Sanskrit is a morphologically rich language, for tasks such as machine translation, question-answering, semantic

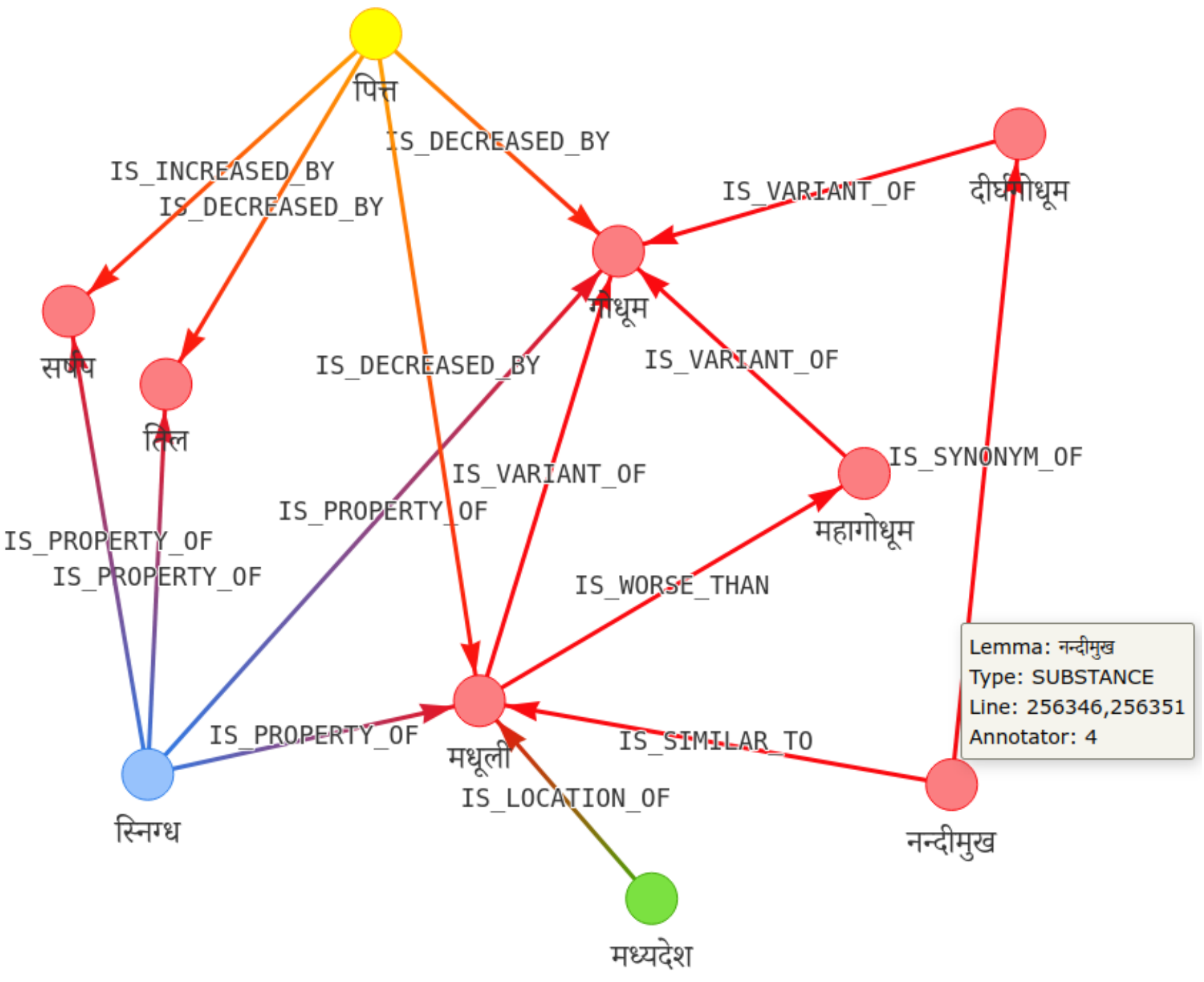

Figure 1: Example of a Knowledge Graph (KG).

labeling, discourse analysis, etc. there are no ready-to-use tools available.

A standard way of capturing knowledge from a text is through the use of *knowledge bases* (KB). It is a form of data repository that stores knowledge in some structured or semi-structured form. A *knowledge graph* (KG) is a particular form of knowledge base that uses the graph data structure to store knowledge. In a KG, nodes represent real-world *entities*, and edges represent *relationships* between these entities. Knowledge about these entities and relationships is typically stored in the form of triplets (*subject, predicate, object*) denoting the relationship predicate a subject has with an object. For example, (Pāṇini, is-author-of, Aṣṭādhyāyī) captures the knowledge nugget 'Pāṇini is the author of Aṣṭādhyāyī'.

An important usage of KGs is automated *question-answering* (QA) where the task is to automatically find answers to questions posed in a natural language. It is an important high-level task in the fields of Information Retrieval (IR) and Natural Language Processing (NLP). Questions can be either from a specific closed domain (such as, say, manuals of certain products) or from the open domain (such as what Google and many other search engines attempt to do). Also, they can be factoid-based (phrase-based or objective) or descriptive (subjective, such as why questions). Since its introduction by Voorhees (1999), one of the main approaches for the question-answering task has been through use of knowledge bases (Hirschman and Gaizauskas, 2001; Kiyota et al., 2002; Yih et al., 2015).

Figure 1 shows an example of (a snippet of) a knowledge graph. The triplets are depicted visually. It contains several nodes (entities) such as madhūlī (मधूली), nandīmukha (नन्दीमुख), pitta (पित्त), snigdha (स्निग्ध), etc. and edges (relationships) including 'is Decreased by', 'is Property of', 'is Variant of', etc. The graph also shows properties associated with the entities and the relationships.

Given a corpus of text, there are many automated ways of constructing knowledge bases from it (Dong et al., 2014; Pujara and Singh, 2018; Mitchell et al., 2018; Wu et al., 2019). These attempts are fairly successful for languages such as English where the state-of-the-art in NLP tools is more advanced. However, due to paucity of such tools in Sanskrit, automated construction of knowledge bases in Sanskrit, to the best of our understanding and knowledge, is only moderately successful. A notable effort is by Terdalkar and Bhattacharya (2019) who attempted to automatically extract all human relationships from Itihāsa texts (Rāmāyaṇa and Mahābhārata) and synonym relationships from an Ayurvedic text (Bhāvaprakāśa). They reported that for an objective natural language query, the correct answer was present in the reported set of answers 50% of the times. They, however, do not report how accurately a triplet is automatically extracted, due to the lack of ground truth for the evaluation.

A more viable and accurate alternative of constructing knowledge bases is through the route of human annotation. *Annotation* of a corpus is the process of highlighting and/or extracting objective information from it. In addition to information extracted from a corpus, the knowledge base may use information that is not directly mentioned in the corpus, such as world knowledge (for example, a person is a living being) or an ontology or a fact specific to the domain of the corpus. Human annotators are typically aware of the domain; although, depending on the task, they need not always be experts in the subject. For example, vāta (वात) has a general meaning as 'wind', but in Ayurvedic context, it refers to the tridoṣa (त्रिदोष) by the name of vāta. This is not directly mentioned in every Ayurvedic text, but any domain expert is aware of this fact.

In this work, we follow the human annotation process of creating a knowledge graph. We choose the chapter Dhānyavarga (धान्यवर्ग) from the Bhāvaprakāśanighaṇṭu (भावप्रकाशनिघण्टु) portion from the Ayurvedic text Bhāvaprakāśa (भावप्रकाश) as the corpus. Bhāvaprakāśa is one of the most prominent texts in Ayurveda, which is an important medical system developed in ancient India and is still in practice. A nighaṇṭu (nighaṇṭu) in the Indic knowledge system is a list of words, grouped into semantic and thematic categories and accompanied by relevant information about these words such as meanings, explanations or other annotations. It is analogous to a glossary in purpose, but differs in structure. In particular, the Bhāvaprakāśanighaṇṭu text, like most of Sanskrit literature, is in padya (verse) form. The text, while loosely following a theme or a structure, is still free flowing. Sanskrit literature contains a large number of such nighaṇṭu texts either as stand-alone books or as parts of other books.

The nighaṇṭu texts, owing to their partial structure are, therefore, amenable to construction of knowledge bases using human annotation. Further, since they contain a wealth of information, they are important resources for building knowledge bases that can be automatically questioned. A benefit of the presence of structure in nighaṇṭu texts is that annotators need not be domain experts as long as the structure is clear.

## 1.1 Contributions

The contributions of this paper are three-fold.

First, we describe a process of constructing a knowledge graph (KG) through manual annotation. This helps to capture the *semantic information* present in the text that is extremely difficult to do otherwise using automated language and text processing methods. The proposed annotation process also enables capturing relationships with entities that are not named directly in the text. We further discuss the curation process and the optimizations performed during the process of knowledge graph creation from the perspective of querying efficiency.

Second, through careful examination of the different types of entities and relationships mentioned in Bhāvaprakāśanighaṇṭu, we create a suitable ontology for annotating the text. We believe that this can be a good starting point for building an ontology for other Ayurvedic texts, and in particular, glossaries.

Third, we annotate one complete chapter from the text (Dhānyavarga), and create a KG from the annotations. For this purpose, we deploy a customized instance of *Sangrahaka*, an annotation and querying framework developed by us previously (Terdalkar and Bhattacharya, 2021). We also create 31 query templates in English and Sanskrit to feed into the templatized querying interface, that aids users in finding answers for objective questions related to the corpus.

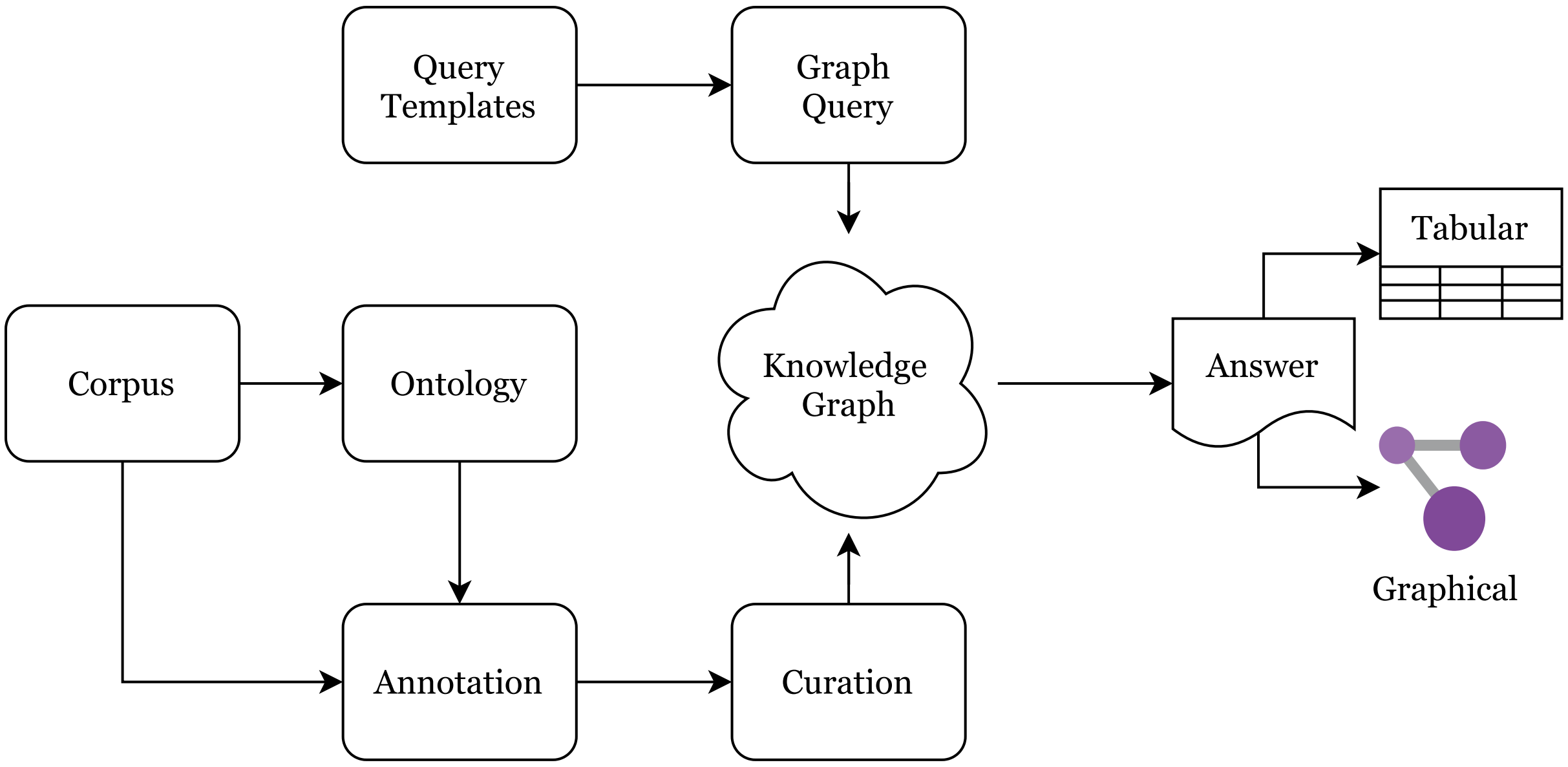

Figure 2: Workflow of semantic annotation for KG construction and querying

The system and the dataset can be accessed at `https://sanskrit.iitk.ac.in/ayurveda/`[1].

### 1.2 Outline

Figure 2 shows the workflow of our proposed method. The first step after inspection of the corpus is of *ontology* creation. After creating a relevant ontology, i.e., specifying what kinds of relationships and entity types are there in the corpus, annotation is performed. Using the entities and relationships captured through annotation, a knowledge graph is constructed. The knowledge graph can be queried with the help of query templates to retrieve answers for templatized natural language questions. Answers are presented in both tabular and graphical formats.

The rest of the paper is organized as follows. §2 motivates the problem of creating a knowledge graph using manual annotations. §3 describes the annotation and curation process along with the construction of knowledge graph. §4 explains the querying mechanism. We conclude in §5 and discuss future directions.

## 2 Motivation for Manual Annotation

To the best of our knowledge, the state-of-the-art in Sanskrit NLP and IR is not advanced enough for automatic construction of knowledge bases from text. One of the first efforts towards automatic construction of knowledge graphs from Sanskrit text was made by Terdalkar and Bhattacharya (2019). The framework described does not yield results comparable to state-of-the-art models for English due to errors in various stages of the construction pipeline. As mentioned earlier, the success rate for even single relationships was not very high.

In this section, we discuss the issues with the Sanskrit state-of-the-art linguistic tools and the need for manual annotation for a *semantic* task such as automatic creation of knowledge graphs.

### 2.1 Word Segmentation

Sanskrit texts make heavy use of compound words in the form of sandhi and samāsa. *Word segmentation*, that splits a given compound word into its constituents is, therefore, an important need in Sanskrit. Notable works in this area are *The Sanskrit Heritage Platform* (SHP) (Huet, 2009; Goyal et al., 2012), *Sanskrit Sandhi and Compound Splitter* (SSCS) (Hellwig and Nehrdich, 2018; Krishna et al., 2016; Krishna et al., 2021).

[1]Please create an account and contact authors requesting access to annotation or querying interface.

Treating the segmentation task as splitting both sandhi and samāsa together, while useful, does not fit well in the pipeline described by Terdalkar and Bhattacharya (2019), where the split output is then passed to a morphological analyzer. An example is splitting of the word maharṣi as mahat + ṛṣiḥ, which while correct as a samāsa-split, if passed to a morphological analyzer as two separate words, produces an analysis[2] of the word mahat independently that does not fit the semantics of the word or the context. Terdalkar and Bhattacharya (2019) applied *SSCS* followed by *SHP*. This results in a word such as rāmalakṣmaṇau getting split into two words rāma and lakṣmaṇau due to *SSCS* resulting in the word rāma getting assigned the vocative case by *SHP*. They used a heuristic to resolve these errors, where the grammatical analysis of the second word was copied to the first word as well. However, this heuristic would change the semantics of the word rāmalakṣmaṇau.

### 2.2 Morphological Analysis

Sanskrit is a highly inflectional language. In Sanskrit, words are categorized as subanta (noun-like) and tiṅanta (verb-like). *Morphological analysis* is the task of identifying the *stem* (prātipadika or dhātu) of the given word form, along with other relevant linguistic information. Notable works in this area are by Goyal et al. (2012) and Kulkarni (2016). These tools perform the best when the input given is without sandhi. If, however, the input also contains splits of samāsa as generated by tools described in the previous section (§2.1), the morphological analyzers treat it as a separate word, resulting in an analysis of the word that may be correct on the syntactic level, but not so in the context of the sentence.

### 2.3 Other Linguistic Tasks

A dependency parser for Sanskrit from *Samsaadhanii* (Kulkarni, 2016) expects the sentences to be in an anvaya order (prose form). Further, it is based on a fixed vocabulary and, therefore, when inflected forms of words from outside the vocabulary are encountered, it fails to parse the sentence. For example, a word śālidhānya is not present in the vocabulary, so a sentence containing that word does not get parsed successfully.

Krishna et al. (2021) in their recent work claim to be able to perform poetry-to-prose linearization and dependency parsing. However, we have not been able to obtain the source code or a functional interface to evaluate it for our data (we contacted the authors).

Another hurdle in the poetry-to-prose linearization is that the sentence boundaries are often not clearly marked. In general, a semantically complete sentence may span over multiple verses. On the other hand, at times a verse may contain multiple sentences as well. This can be seen in the sample of 10 verses given in Appendix A. Thus, extracting sentences with proper sentence boundaries is also a difficult task.

### 2.4 Semantic Information Extraction

Extracting the semantics of a sentence is a very important step in the construction of a knowledge graph. Automatic KG construction frameworks for English such as (Auer et al., 2007; Suchanek et al., 2007) extract semantic information from various information sources including Wikipedia articles and info-boxes. One of the challenges faced in this task is that the same concept can be expressed in English in numerous ways, such as "birthplace" or "place of birth". The issue of expressing a concept in more than one ways is extremely significant and much more severe for Sanskrit due to its semantic richness. In particular, the processes of samāsa and sandhi create long and semantically rich words.

Table 1 highlights this phenomenon. The first column contains the concept while the second column enlists the words used in Dhānyavarga to express that concept. A "concept" captures the semantics of a word or a phrase.

It can be noted that even in the span of 90 verses, there are more than 10 different ways used to express the same concept '(a substance) decreases vāta'. In addition to that, the word vāta itself can be part of another compound, coupled with other words as can be seen in the example 'decreases vāta and pitta'. There are more than 5 usages of this complex concept, which is a superset of the earlier concept. Moreover, these are not the only ways in which the concept of increasing vāta is expressed.

[2]Analysis: mahat (`n. sg. acc. | n. sg. nom.`)

| Concept | Words or Phrases |
|---|---|
| increases bala | balya, balada, balāvaha, balaprada, balakara, balakṛt |
| increase vāta | vātala, vātakṛt, vātakara, vātajanaka, vātajananī, vātātikopana, vātaprakopaṇa, vātavardhana, vātakopana |
| decreases pitta | pittaghna, pittapraṇāśana, pittapraśamana, pittahara, pittaghnī, pittajit, pittāpaha, pittajit, pittahṛt, pittanut, pittavināśinī |
| decreases vāta and pitta | vātapittaghna, pittavātaghna, pittavātavibandhakṛt, vātapitta-hara, vātapittahṛt |

Table 1: Semantic variations due to richness of Sanskrit through examples from Dhānyavarga.

There are numerous other words that can combine with the word vāta in the form of samāsa to indicate the concept of decrement for multiple entities at the same time. Moreover, in such cases, where a samāsa is used, the order of vāta and pitta could be reversed as well. Further, this list for a particular concept is not exhaustive for Sanskrit, and there are practically endless possible ways to denote the same concept.

One can observe that there are some common suffixes used in similar concepts. However, firstly, there is no exhaustive list of suffixes available associated with a particular concept. Second, the suffixes have different concepts in different contexts. For example, the suffix -ghna (-घ्न) in the context of Ayurveda, or specifically of tridoṣa, means 'to decrease', e.g., pittaghna (pittaghna). The same suffix in the context of a person may mean 'to kill', e.g., śatrughna (शत्रुघ्न) – one who kills his enemies (śatru).

Thus, using a fixed set of suffixes may not be a feasible solution.

To the best of our knowledge, there is no existing system for Sanskrit that can extract such semantic information in either a generic sense or in a specific context. Amarakoṣa Knowledge Net (Nair and Kulkarni, 2010) and Sanskrit WordNet (Kulkarni et al., 2010) are also limited in their scope. For example, none of the words listed in the Table 1 to express the concept of 'increasing bala' can be found in either of these two resources.

### 2.5 Need for Annotation

Issue of compounding errors is relevant to any NLP pipeline, where individual parts of the pipeline have their own error rates. The success rate of the entire pipeline, being a multiplicative factor of the individual success rates (since all the parts have to be accurate for the entire task to be accurate), is significantly lower. Thus, the pipeline for the automated question-answering task that requires modules such as word segmentation, morphological analysis, part-of-speech tagging, dependency parsing, etc. has a very low accuracy. Further, the lack of semantic analysis tools or systems is a major hurdle in semantic tasks such as construction of knowledge graphs. Thus, even if the accuracy of the individual parts are improved significantly, the final semantic labeling remains a bottleneck.

We highlight this fact by taking an example of the first $\sim 10\%$ of the Dhānyavarga, i.e., 10 verses corresponding to 21 lines. We have manually segmented the words in these lines and also converted the sentences to anvaya order. The first 10 verses correspond to a total of 14 prose sentences. The original text in verse format, in the sandhi-split format, and in anvaya format, is given in Appendix A.

There are a total of 35 occurrences of sandhi and 50 occurrences of samāsa in the text. *SSCS* is able to identify 34 of the sandhi (with an accuracy of 0.97) and 34 occurrences of samāsa correctly (with an accuracy of 0.68). However, the tool does not differentiate a sandhi from a samāsa. Therefore, when passed to the *SHP* it is likely to obtain incorrect analysis.

A single word-form in Sanskrit can have numerous valid morphological analyses. If there are $N$ words in a sentence, and every word has $a_i$ analyses possible, then there are $\Pi_{i=1}^{N} a_i$ possible combinations for the correct analysis of the sentence. *SHP* and *Samsaadhanii* both rank these solutions based on various linguistic features, and after pruning the unlikely ones, present the feasible solutions. For automatic processing pipelines, a particular choice of the solution is required, and the solution presented as the *best* by the tools, i.e., the first solution, is a natural choice. Thus, we present the evaluation by choosing the first solution.

We pass the manually created sandhi-split corpus through *SHP* for morphological analysis.[3] There are an average of 9 solutions per line (ranging from 0 to 72) reported. We evaluate based on the first reported solution.

There are 103 words, after manually splitting sandhi. *SHP* could not analyze 21 words, and wrongly analyzed 14 words, resulting in an overall accuracy of 0.66. Further, *SHP* split 34 words, of which 8 were incorrect splits, resulting in an accuracy of 0.76 for samāsa-split.

Additionally, we pass the anvaya-order sentences to the dependency parser tool by Kulkarni (2016). We also manually add missing verbs (adhyāhāra forms such as asti, santi, etc.) due to that being a requirement of the parser. Without samāsa split markers, the dependency parser manages to parse only 1 out of 14 sentences, while with the samāsa markers, it can parse 6 out of 14. Out of the 6 sentences that produce a dependency parse tree, 4 are simple 3-word sentences (sentences 2, 3, 4, 6 in Appendix A). In the other 2 instances (sentences 7, 10), errors were found in the dependency parse trees.

For example, consider a line from śloka 2:

> Sanskrit: ⟨ कङ्ग्वादिकं क्षुद्रधान्यं ⟩ ⟨ तृणधान्यञ्च तत्स्मृतम् ⟩
> IAST: ⟨ kaṅgvādikaṃ kṣudradhānyaṃ ⟩ ⟨ tṛṇadhānyañca tatsmṛtam ⟩
> Meaning: kaṅgu etc. are types of kṣudradhānya. It (kṣudradhānya) is also called tṛṇadhānya.
> Anvaya: क्षुद्रधान्यम् कङ्ग्वादिकम् (अस्ति)। तत् तृणधान्यम् च स्मृतम् (अस्ति)।

There are two sentences in this line, as can be seen by the boundary markers and anvaya text. Figure 3 shows the dependency parse for these two sentences. The dependency parse for the first dependency tree is correct. However, even for the sentences that get a correct dependency parse, the dependency relations we get are kartā and kartṛsamānādhikaraṇa which, while grammatically correct, still do not help in capturing the *semantic* concept of the sentence that kaṅgu is a *type* of kṣudradhānya. Also, tat in the second sentence is an *anaphora* of kṣudradhānya from the first sentence. Thus, the intended relation that tṛṇadhānya is a synonym of kṣudradhānya cannot be extracted without a module for *anaphora resolution.* Yet again, to the best of our knowledge, there is no such co-reference resolution system for Sanskrit.

More importantly, no existing tool has a capability of performing *semantic* tasks, which are a requirement for knowledge extraction. Manual annotation, therefore, is the only way to capture the semantic relations. In addition, it bypasses the entire NLP pipeline and, thus, has a high potential for creating a question-answering system that is much more accurate and reliable than a system based on automatically created knowledge graphs.

Another prevalent issue is the lack of datasets for training and evaluation of tasks such as question-answering or creation of knowledge bases. Creation of knowledge bases through manual annotation is, thus, of utmost importance both for the actual task of question-answering and for further research in the field, including automated knowledge base construction since these may act as ground-truth benchmark datasets for evaluation of future automated tools.

---

[3] We keep a timeout of 60 seconds, within which if the analysis is not found, we report the analysis as missing, i.e., 0 solutions for that line.

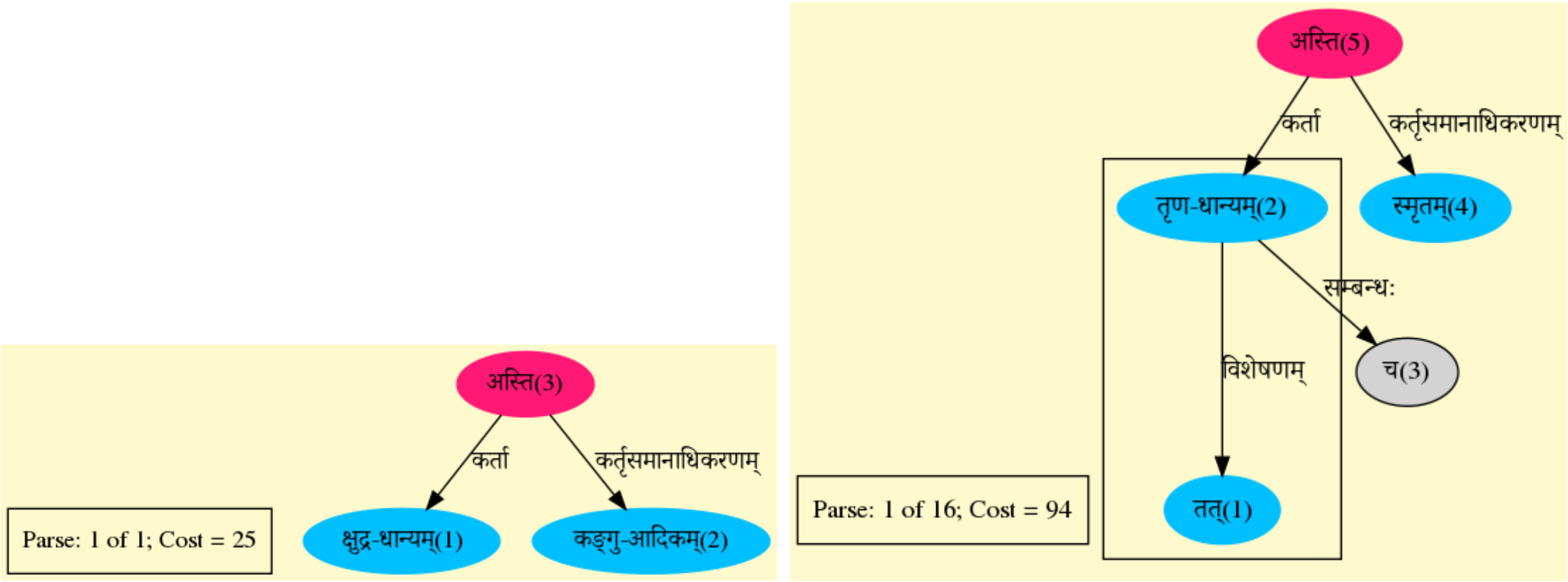

Figure 3: Dependency parse trees for sentences from śloka 2.

Bhavaprakasha Nighantu - धान्यवर्गः

Search

| | LINE | TEXT | SPLIT | ? |
|---|---|---|---|---|
| ⊙ | 256381 | मसूरो मधुरः पाके संग्राही शीतलो लघुः | मसूरः मधुरः पाके संग्राही शीतलः लघुः | ✔ |

| WORD | मसूरः | मधुरः | पाके | सङ्ग्राही | शीतलः | लघुः |
|---|---|---|---|---|---|---|
| ROOT | मसूर | मधुर | पाक | सङ्ग्राहिन् | शीतल | लघु |
| GENDER | m. | m. | m. | m. | m. | m. |
| CASE | 1 | 1 | 7 | 1 | 1 | 1 |
| NUMBER | sg. | sg. | sg. | sg. | sg. | sg. |

Figure 4: Sample text from Dhānyavarga with linguistic information

## 3 Annotation Process

Annotation has been done with the purpose of building a knowledge graph (KG). We fix the unit for annotation to be a line from a verse (śloka). We collect annotations of two types — *entity* and *relation* — described in detail in §3.3 and §3.4 respectively.

The corpus interface from *Sangrahaka* is capable of displaying extra information about each line. We use this feature to display word segmentation and morphological analysis of the text produced by *SSCS* and *SHP*, which can potentially help the annotators. Figure 4 shows a sample text from Dhānyavarga with linguistic information.

An annotator goes through the lines assigned to her and for each line, identifies the entities as well as the relationships between the entities appearing in it.

### 3.1 Corpus

Bhāvaprakāśanighaṇṭu is the nighaṇṭu portion of Bhāvaprakāśa. It contains a list and description of various medicinally relevant plants, flowers, fruits, animals, grains, animal products, metals, prepared substances, etc. These are divided into 23 chapters called vargas.

A general structure followed in the Bhāvaprakāśanighaṇṭu is as follows,

| | |
|---|---|
| Entities (25) | Substance, Part of a Substance, Compound Substance, Prepared Substance, Collection of Substances, Tridoṣa, Property, Effect, Disease, Symptom, Product/Waste of Human Body, Part of Human Body, Person, Animal, Plant, Source, Animal Source, Plant Source, Quantity, Method or Preparation, Usage, Location, Time, Season, Others |
| Relationships (29) | is Synonym of, is Type of, is Variant of, is Property of, is (Not) Property of, is Similar to, is Better/Larger/Greater than, is Worse/Smaller/Lesser than, is Newer than, is Older than, is Best/Largest/Greatest among, is Medium among, is Worst/Smallest/Least among, is Ingredient of, is Part of, is (Not) Part of, is Disease of, is Caused by, is (Not) Caused by, is Benefited by, is Harmed by, is Produced by, is Removed/Cured by, is Increased by, is Decreased/Reduced by, is Preparation of, is (Absence/Lack of) Preparation of, is Location of, is Time of |

Table 2: Entity and relationship types in our ontology

- Substances are semantically grouped in chapters. For example, all grains appear in the chapter Dhānyavarga.
- Each chapter contains several virtual sections pertaining to a single substance. Only when a substance has been described in entirety, discussion about another substance starts.[4]
- Each section about a substance has the following information:
  - Synonyms, if any, of the substance
  - Properties, e.g., color, smell, texture
  - Effects, e.g., effect on tridoṣa (vāta, pitta, and kapha)
  - Symptoms and diseases treated or cured by the substance
  - Variants, if any, of the substance
  - Properties of each variant, and their distinguishing characteristics
  - Comparison between the variants, if possible
  - Time and location where the substance is found or grown, if relevant
- The order of information components about a substance within a section may vary.

The entire Bhāvaprakāśanighaṇṭu contains 2087 verses, corresponding to 4201 lines. We have chosen Dhānyavarga, a chapter about grains, which has a wide variety of entity types and relations. It contains 90 verses, corresponding to 183 lines.

### 3.2 Ontology

We have created an ontology consisting of 25 entity types and 29 relationship types by carefully going through multiple chapters from the Bhāvaprakāśanighaṇṭu including Dhānyavarga. An exhaustive list of them is mentioned in Table 2.

The decision to add a certain entity type or relation type is made based on the importance of the concept, frequency of its occurrence, and nature of frequently asked questions.

For example, the concept of vāta, pitta and kapha, collectively referred to as tridoṣa or fundamental elements (humors) of the body, is central to Ayurveda. Consequently, queries such as “What effect does a substance X have on a (one of the) tridoṣa?” is one of the most fundamental and common information requirement about the substance. Therefore, we have decided to create a category “Tridoṣa” only for these three entities. Therefore, any occurrence of these words or their synonyms, e.g., śleṣma, a synonym of a kapha, results in the creation of an entity of type Tridoṣa.

[4]There is, however, no indication in the original text that a section/substance has ended and a new one has started. It must be inferred by reading the text.

Similarly, the type of effect any substance has on each of the tridoṣa is either an increment or decrement. Therefore, we have identified relations is Increased by and is Decreased by.

After the ontology has been finalized, the next step is annotation.

### 3.3 Entity Annotation

Entities correspond to nodes in the knowledge graph. When a word that represents an entity is encountered, its *lemma* (prātipadika) and the entity type it belongs to are identified, and the entity is marked.

As an example, consider the following line from śloka-31 of Dhānyavarga:

> Devanagari: गोधूमः सुमनोऽपि स्यात्त्रिविधः स च कीर्त्तितः।
> IAST: godhūmaḥ sumano'pi syāttrividhaḥ sa ca kīrttitaḥ.
> Meaning: Godhūma (wheat) is also called Sumana, and it is said to be of three kinds.

Here, there are two entities, godhūma and sumana, both of type “Substance”. An entity needs to be added explicitly only the first time it is encountered.

In a case where a samāsa is used to indicate an effect on an entity, and the relation fits one of the relationship types, a relevant word (pada) from the segmentation (vigraha) of samāsa is used. For example, consider the following line from śloka-33:

> Devanagari: गोधूमः मधुरः शीतो वातपित्तहरो गुरुः।
> IAST: godhūmaḥ madhuraḥ śīto vātapittaharo guruḥ.
> Meaning: Godhūma is sweet, cold, hard to digest and removes (decreases) vāta and pitta.

Here, vātapittaharaḥ is a single word, which uses samāsa, to indicate that vāta and pitta are reduced by godhūma. Therefore, vātapittahara will not be added as an entity; instead the entities vāta and pitta are recognized.

### 3.4 Relation Annotation

Relations correspond to edges in the knowledge graph. A relation, which fits one of the relationship types from the ontology, is identified by interpreting the śloka. *Subject* and *Object* for this relation are then identified. Relations, where extra semantic information is known, such as madhura is known to be a rasa, are endowed with that extra information.

Consider the two examples of lines mentioned in the previous section (§3.3), śloka-31 and śloka-33. Following relations are added based on these two lines:

sumana $\vdash$ is Synonym of $\rightarrow$ godhūma
madhura $\vdash$ is (rasa) Property of $\rightarrow$ godhūma
śīta $\vdash$ is Property of $\rightarrow$ godhūma
vāta $\vdash$ is Decreased by $\rightarrow$ godhūma
pitta $\vdash$ is Decreased by $\rightarrow$ godhūma
guru $\vdash$ is Property of $\rightarrow$ godhūma

It should be noted that neither the subject nor the object may be present as words in the line that mentions a relationship about it. Consider, for example, the next line of the śloka-33:

> Devanagari: कफशुक्रप्रदो बल्यः स्निग्धः सन्धानकृत्सरः।
> IAST: kaphaśukraprado balyaḥ snigdhaḥ sandhānakṛtsaraḥ.
> Meaning: (Godhūma) increases kapha, śukra, bala, is snigdha, sandhānakṛt (helps in joining broken bones) and laxative.

Here, the description of properties of godhūma (from previous line) is continued. Therefore, one of the relations added is

kapha ⊢ is Increased by → godhūma

This relation has godhūma as *Object*, although it is not present in the line itself.

### 3.5 Unnamed Entities

On occasions, it may happen that an entity is referenced by its properties only, and it is not named at all in the text. Consider the following line from śloka-39:

Devanagari: मुद्गो बहुविधः श्यामो हरितः पीतकस्तथा।
IAST: mudgo bahuvidhaḥ śyāmo haritaḥ pītakastathā.
Meaning: Mudga is of various types – black, green, and yellow.

Thus, there are three colored variants of the substance mudga, but they are not named. In such a case, we create three *unnamed entities* (denoted by `X`-prefixed nodes) with entity type “Substance”, same as that of mudga to refer to the three varieties. Each of these entities is given a unique identifier. The unique identifier is a combination of the unnamed entity number and the line number it occurs in. Thus, if the line number is 256358, the black variant is given the identifier `X1-256358`. Similarly, the green variant is identified as `X2-256358` while the yellow variant is identified as `X3-256358`.

To describe these variants, three relations are added as well:

śyāma ⊢ is (varṇa) Property of → `X1-256358`
harita ⊢ is (varṇa) Property of → `X2-256358`
pīta ⊢ is (varṇa) Property of → `X3-256358`

The utility of such annotations becomes clear when these unnamed entities are later referred to in another line or another verse.

The next line of śloka-39 reads

Devanagari: श्वेतो रक्तश्च तेषान्तु पूर्वः पूर्वो लघुः स्मृतः ॥३९॥
IAST: śveto raktaśca teṣāntu pūrvaḥ pūrvo laghuḥ smṛtaḥ. ||39||
Meaning: … white and red. Among them, each is successively easier to digest.

The word teṣām here refers to the five varieties of mudga, and gives a relation between them. So, we get two new unnamed entities in this line, `X1-256359` and `X2-256359` (note how `X1` and `X2` are re-used but with different line numbers).

We also get a total of four new relations to capture the successive ease in digestion properties:

`X1-256358` ⊢ is Better (in property laghu) than → `X2-256358`
`X2-256358` ⊢ is Better (in property laghu) than → `X3-256358`
`X3-256358` ⊢ is Better (in property laghu) than → `X1-256359`
`X1-256359` ⊢ is Better (in property laghu) than → `X2-256359`

For the purpose of querying, the anonymous nodes are treated like any other node.

### 3.6 Auto-complete Suggestions

We have enhanced the annotation interface from *Sangrahaka* to improve user experience with Sanskrit text by adding transliteration-based suggestions. There are numerous standard schemes for *Devanagari* transliteration[5]. Whenever, a *Devanagari* entity is annotated, we transliterate it using `indic-transliteration` package (Sanskrit programmers, 2021) into all the available schemes. We maintain an index with all the transliterations. Now, when a user enters any text, we query our index and return all suggestions that match with the lower-cased version of the user text. For example, consider a word in Devanagari ‘माष’, which transliterates into ‘mASa’ (HK), ‘mASha’ (ITRANS), ‘māṣa’ (IAST), ‘maa.sa’ (Velthuis), ‘mARa’ (WX) and ‘mAza’ (SLP1). Now, a user may enter at least 3 starting characters from any of the scheme, e.g., ‘mas’, ‘maa’, 'maz', 'mar' etc. and the Devanagari word 'माष' will be suggested. The index is maintained globally. So, once an entity is entered by any annotator, the completions for that entity become available to all annotators.

[5]`https://en.wikipedia.org/wiki/Devanagari_transliteration`

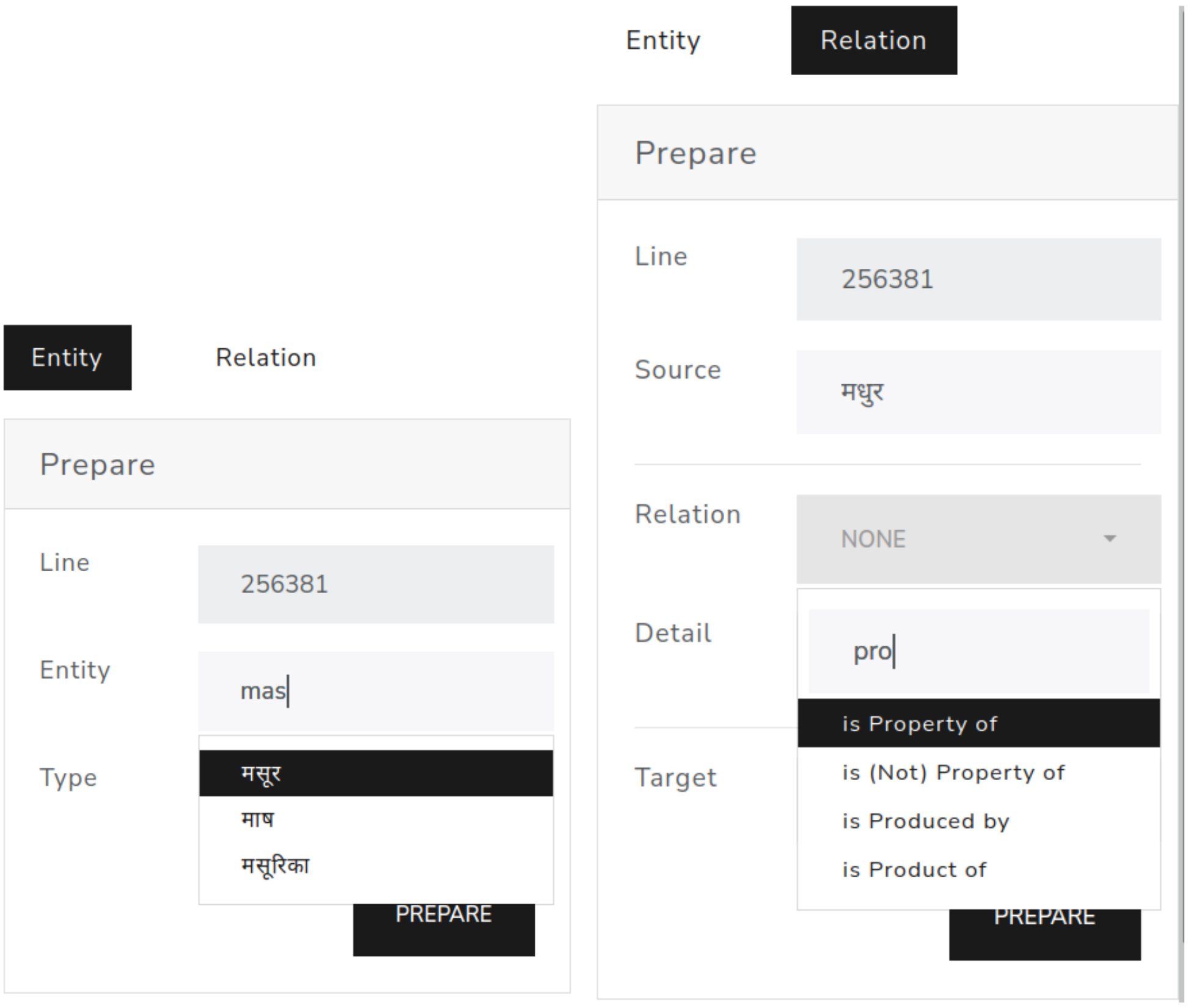

Figure 5: Modified annotation interface with multi-transliteration-based suggestions

These suggestions are enabled to all text input fields, namely, entity annotation, relationship annotations and querying interface.

Figure 5 shows the modified annotation interface with auto-complete suggestions.

## 3.7 Curation

After the annotation step, before construction of the knowledge graph, a thorough curation step is required to resolve errors or inconsistencies that may inadvertently creep up during the annotation process.

### 3.7.1 Equivalent Entities

The linguistic information that we have added with the corpus is supposed to serve as a guideline for the annotation. However, since this information is generated using automated tools, there might be errors. For example, the word grāhī (ग्राही) refers to substances that have a property of absorbing liquid and increasing digestive power. The reported prātipadika of this word, automatically generated by *SHP*, is grāha (ग्राह) instead of grāhin. An annotator, by oversight, may mark the incorrect lemma. Additionally, for substance names in feminine gender, which also have this property, an adjective grāhiṇī (ग्राहिणी) is used. The correct prātipadika in this case would be grāhiṇī. The node refers to the same property. So, semantically they are equivalent to each other, and ideally should be captured using a single name. These instances are common for properties of substances.

To address this issue, we add a relation is Synonym of between these entities. This, in conjunction with the optimization mechanism described in §3.8, tackles the issue of equivalent entities.

### 3.7.2 Inconsistent Node Categories

There may be differences of opinions between annotators regarding which category a particular node should belong to. For example, an entity jvara (ज्वर) refers to fever. This entity was marked as a *Symptom* by some annotators and as a *Disease* by the others. Such cases were resolved through discussion among the curators.

### 3.7.3 Missing Node Categories

The framework allows entities to be mentioned in the relationships without being added as entities. While care was taken to always add entities before marking relationships involving those entities, there may still be instances of human error, where an annotator may forget to mark an entity. We created a set of inference rules to infer as many instances of such occurrences as possible. For example, if an entity is marked as a *source* of the relation is Property of, without having been added as an entity, we can automatically create that entity by assigning the category '*Property*' to it.

## 3.8 Symmetric Relationships

The relation is Synonym of is symmetric, i.e., if `A` is a synonym of `B`, then, by definition, `B` is also a synonym of `A`. A query can be made using any of the synonyms, and the system should still be able to return the correct answer.

Suppose, $S_1, S_2 \dots, S_N$ are $N$ synonyms of a substance. If the synonym group is completely captured, then a user should be able to query using any synonym and still get the desired result. Properties of the substance corresponding to this synonym group can also be scattered across $S_i$'s. Say, there are $M$ properties $P_1, P_2, \dots, P_M$, and some of the relations are $P_1$ is Property of $S_1$, $P_2$ is Property of $S_4$, $P_3$ is Property of $S_3$, and so on. Now, if we want to query whether substance $S_2$ has property $P_1$ but we search using the name of the substance as $S_2$, a direct query will not work, as there is no direct relation between the nodes $S_2$ and $P_1$. For the correct answer, we would have to find every synonym of $S_2$ and check if any of them has the property $P_1$. This requires a *path query*. A path query involving $N$ synonyms may require as many as $N-1$ edge traversals. Path queries are NP-hard (Mendelzon and Wood, 1995) and are, therefore, computationally expensive.

For example, rājikā, kṣava, kṣutābhijanaka, kṛṣṇīkā, kṛṣṇasarṣapa, rājī, kṣujjanikā, āsurī, tīkṣṇagandhā, cīnāka are all names of the same substance. A relation is added as follows,

uṣṇa $\vdash$ is Property of $\rightarrow$ rājikā

Now, suppose that we want to query for a property of the substance kṣava, which while referring to the same entity as rājikā, does not have a property edge incident upon it.

We, therefore, are forced to use a path query, and the query has to explore all the synonym paths from kṣava to find out if kṣava itself or one of its synonyms has any property edge. The number of such paths can be impractically large, especially for large knowledge graphs.

We perform a simple optimization heuristic to tackle this issue. We first identify a synonym $S_K$ among all the synonyms having the highest degree, i.e., $K \in \{1, .., N\}$, such that $K = \underset{i}{\text{argmax}}\ degree(S_i)$. We treat this as the *canonical name* for that synonym group, and we add a relation is Synonym of from every $S_i, i \neq K$ to $S_K$. Further, we transfer all the edges (other than the `is Synonym of` edge) from every $S_i, i \neq K$ to $S_K$. In other words, if $S_i$ was connected to a node $V$ by a relation $R$, after optimization, $S_K$ will be connected to node $V$ by relation $R$. Now, every synonym has a direct edge to the canonical name, with all the properties getting attached to the canonical name only. Thus, a query on any synonym has to traverse *at most 1 edge* before reaching the desired node.

At the end of curation and optimization steps, there were 410 nodes and 764 relationships that constitute our knowledge graph.

# 4 Querying

Although the ideal way of question-answering is by posing queries in natural language, unfortunately, the state-of-the-art in Sanskrit NLP tools does not allow that. Hence, to simulate natural language queries, we use query templates.

The annotation and querying platform that we use, *Sangrahaka*, uses *Neo4j* graph database[6] for the purpose of storing and querying the knowledge graph. *Cypher*[7] is *Neo4j*'s graph query language inspired by SQL, but optimized for graph querying and it makes use of intuitive ASCII-art syntax for querying. The platform utilizes the power of *Cypher* for connecting to the graph database. Natural language queries are simulated using query templates.

## 4.1 Query Templates

A query template consists of a set of natural language templates and an equivalent graph query template. Each of these templates contain *placeholders*. Values of these placeholders can be filled by choosing the required entity, entity type or relation, to convert the query template into a valid natural language query. The same replacement in the graph query template yields a valid graph query which can be directly used to fetch results from the graph database.

For example, consider a sample query template:

- Sanskrit: के पदार्थाः `{0}` इति दोषस्य वर्धनं कुर्वन्ति।
- English: `Which entities increase the dosha {0}?`
- Cypher:

```
MATCH (dosha:TRIDOSHA)-[relation:IS_INCREASED_BY]->(entity)
WHERE dosha.lemma =  "{0}"
RETURN entity
```

The variable `{0}` here is a word representing an entity of type `TRIDOSHA`. The valid values for the variable in this query are vāta or pitta or kapha or one of their synonyms. So, natural language questions such as "Which substances increase kapha?", etc. can be realized using this query template.

In order to increase the number of the questions that can be answered, we have created a set of *generic* queries which help model *any* query up to a single relation. It contains the following three query templates:

- Which entity is related to entity `{0}` by relation `{1}`?
- How is entity `{0}` related to entity `{1}`?
- Show all matches where an entity of type `{0}` has relation `{1}` with an entity of type `{2}`.

We have a total of 31 *natural language query templates* in Sanskrit[8] to represent the most relevant queries. We have classified these templates semantically into 12 categories. Classification helps to locate an intended query template faster. An exhaustive list of these query templates and their categories is in Appendix B.

## 4.2 Query Answers

Result of graph queries are also graphs. The querying interface from *Sangrahaka* consisted of a graphical and a tabular display. Figure 6 shows a sample output using the query interface. In the graph, hovering a node lets the user see the properties associated with the node by hovering over the node in the graph. Each node label is a *lemma* (word-stem) associated with that node. In addition, provenance of the node such as which line from the corpus does that node correspond to, and the identifier of the annotator(s) who added that entity are also mentioned. The nodes are color-coded in such a way that nodes referring to entities of same type get the same color[9].

[6] `https://neo4j.com/`

[7] `https://neo4j.com/developer/cypher/`

[8] We also have their English translated versions in the system.

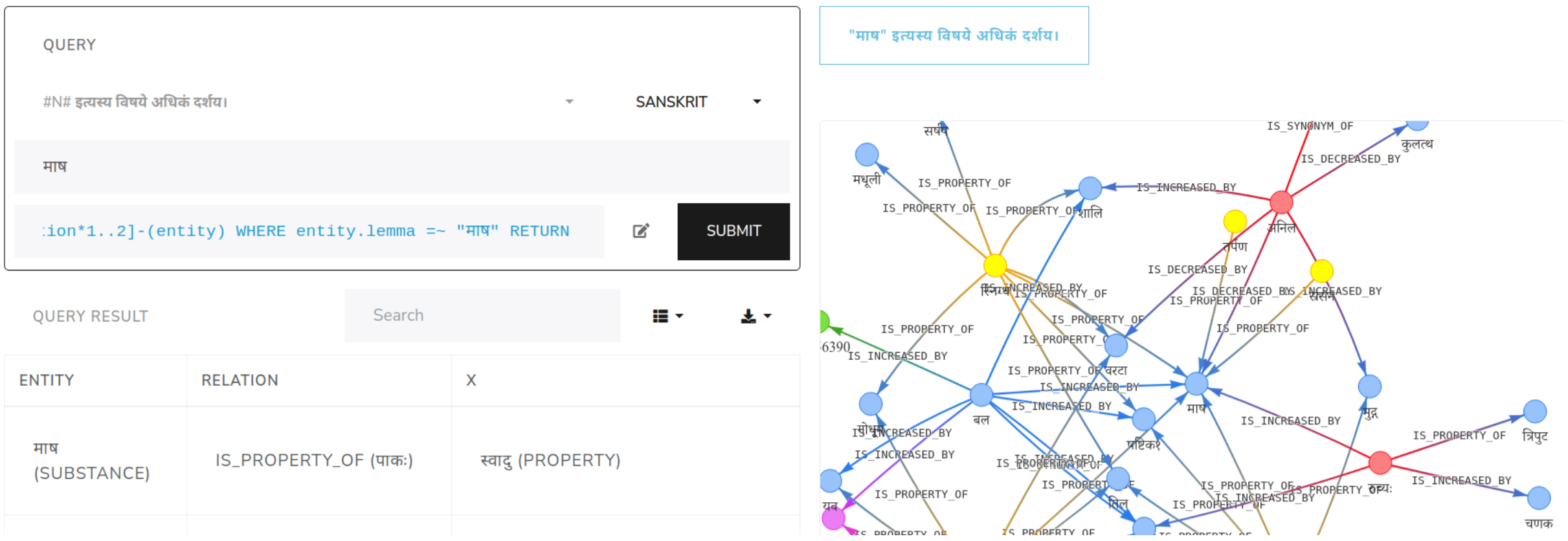

Figure 6: Sample output using query interface featuring Sanskrit query templates

## 5 Conclusions and Future Work

Current state of Sanskrit NLP makes manual annotation a necessity for semantic tasks. We propose a construction of a knowledge graph (KG) through manual annotation process with a special focus on capturing semantic information. We also introduce a mechanism to handle unnamed entities in a knowledge graph.

The deployed instance of the framework used for the purpose of annotation and querying can be accessed at `https://sanskrit.iitk.ac.in/ayurveda/`.

As a proof-of-concept, we have selected a chapter from the nighaṇṭu text Bhāvaprakāśanighaṇṭu, carefully created an ontology, and performed semantic annotation to construct a knowledge graph. Our methodology is extensible to other nighaṇṭu texts.

In future, we plan to complete the annotation of the rest of the Bhāvaprakāśanighaṇṭu. We also plan to explore more classical texts such as Rāmāyaṇa and Mahābhārata for annotating other kinds of relationships.

We also hope that the dataset created in the process will prove useful for further research efforts in the area of NLP in Sanskrit. We make the ontology and the dataset available at `https://sanskrit.iitk.ac.in/ayurveda/dataset/`.

## Acknowledgements

We thank Anil Kumar Gourishetty from IIT Roorkee for his help in connecting us with Sanskrit volunteers.

[9] Colors are not fixed. Thus, the color *yellow* is not indicative of a specific entity type. It only ensures that for a particular query answer, all yellow nodes will have the same entity type.

# A Dhānyavarga Sample

## A.1 Sample of Text

We present here an extract from Dhānyavarga used in §2. Table 3 contains the first 10 verses from Dhānyavarga and a version with sandhi resolved. The sentence boundaries are denoted using '⟨' and '⟩' markers.

| **Original Text** | Sandhi **Split** |
|---|---|
| ⟨ शालिधान्यं व्रीहिधान्यं शूकधान्यं तृतीयकम्<br>शिम्बीधान्यं क्षुद्रधान्यमित्युक्तं धान्यपञ्चकम् १ ⟩ | शालिधान्यम् व्रीहिधान्यम् शूकधान्यम् तृतीयकम्<br>शिम्बीधान्यम् क्षुद्रधान्यम् इति उक्तम् धान्यपञ्चकम् १ |
| ⟨ शालयो रक्तशाल्याद्या ⟩ ⟨ व्रीहयः षष्टिकादयः ⟩<br>⟨ यवादिकं शूकधान्यं ⟩ ⟨ मुद्गाद्यं शिम्बिधान्यकम् ⟩<br>⟨ कङ्ग्वादिकं क्षुद्रधान्यं ⟩ ⟨ तृणधान्यञ्च तत्स्मृतम् २ ⟩ | शालयः रक्तशाल्याद्याः व्रीहयः षष्टिकादयः<br>यवादिकम् शूकधान्यम् मुद्गाद्यम् शिम्बिधान्यकम्<br>कङ्ग्वादिकम् क्षुद्रधान्यम् तृणधान्यम् च तत् स्मृतम् २ |
| ⟨ कण्डनेन विना शुक्ला हैमन्ताः शालयः स्मृताः ३ ⟩ | कण्डनेन विना शुक्लाः हैमन्ताः शालयः स्मृताः ३ |
| ⟨ रक्तशालिः सकलमः पाण्डुकः शकुनाहृतः<br>सुगन्धकः कर्दमको महाशालिश्च दूषकः ४ | रक्तशालिः सकलमः पाण्डुकः शकुनाहृतः<br>सुगन्धकः कर्दमकः महाशालिः च दूषकः ४ |
| पुष्पाण्डकः पुण्डरीकस्तथा महिषमस्तकः<br>दीर्घशूकः काञ्चनको हायनो लोध्रपुष्पकः ५ | पुष्पाण्डकः पुण्डरीकः तथा महिषमस्तकः<br>दीर्घशूकः काञ्चनकः हायनः लोध्रपुष्पकः ५ |
| इत्याद्याः शालयः सन्ति बहवो बहुदेशजाः ⟩<br>⟨ ग्रन्थविस्तरभीतेस्ते समस्ता नात्र भाषिताः ६ ⟩ | इत्याद्याः शालयः सन्ति बहवः बहुदेशजाः<br>ग्रन्थविस्तरभीतेः ते समस्ताः न अत्र भाषिताः ६ |
| ⟨ शालयो मधुराः स्निग्धा बल्या बद्धाल्पवर्चसः<br>कषाया लघवो रुच्याः स्वर्या वृष्याश्च बृंहणाः<br>अल्पानिलकफाः शीताः पित्तघ्ना मूत्रलास्तथा ७ ⟩ | शालयः मधुराः स्निग्धाः बल्याः बद्धाल्पवर्चसः<br>कषायाः लघवः रुच्याः स्वर्याः वृष्याः च बृंहणाः<br>अल्पानिलकफाः शीताः पित्तघ्नाः मूत्रलाः तथा ७ |
| ⟨ शालयो दग्धमृज्जाताः कषाया लघुपाकिनः<br>सृष्टमूत्रपुरीषाश्च रूक्षाः श्लेष्मापकर्षणाः ८ ⟩ | शालयः दग्धमृज्जाताः कषायाः लघुपाकिनः<br>सृष्टमूत्रपुरीषाः च रूक्षाः श्लेष्मापकर्षणाः ८ |
| ⟨ कैदारा वातपित्तघ्ना गुरवः कफशुक्रलाः<br>कषायाश्चाल्पवर्चस्का मेध्याश्चैव बलावहाः ९ ⟩ | कैदाराः वातपित्तघ्नाः गुरवः कफशुक्रलाः<br>कषायाः च अल्पवर्चस्काः मेध्याः च एव बलावहाः ९ |
| ⟨ स्थलजाः स्वादवः पित्तकफघ्ना वातवह्निदाः<br>किञ्चित्तिक्ताः कषायाश्च विपाके कटुका अपि १० ⟩ | स्थलजाः स्वादवः पित्तकफघ्नाः वातवह्निदाः<br>किञ्चिद् तिक्ताः कषायाः च विपाके कटुकाः अपि १० |

Table 3: First 10 verses from Dhānyavarga of Bhāvaprakāśanighaṇṭu

## A.2 Poetry-to-Prose Conversion of Verses from Table 3

We next list the prose version of the verses listed in Table 3 above.

1. शालिधान्यम् व्रीहिधान्यम् तृतीयकम् शूकधान्यम् शिम्बीधान्यम् क्षुद्रधान्यम् इति धान्यपञ्चकम् उक्तम् (अस्ति)।
2. शालयः रक्तशाल्याद्याः (सन्ति)।
3. व्रीहयः षष्टिकादयः (सन्ति)।
4. शूकधान्यम् यवादिकम् (अस्ति)।
5. शिम्बिधान्यकम् मुद्गाद्यम् (अस्ति)।
6. क्षुद्रधान्यम् कङ्ग्वादिकम् (अस्ति)।
7. तत् तृणधान्यम् च स्मृतम् (अस्ति)।

8. कण्डनेन विना शुक्लाः हैमन्ताः (च) शालयः स्मृताः (सन्ति)।
9. शालयः रक्तशालिः सकलमः पाण्डुकः शकुनाहृतः सुगन्धकः कर्दमकः महाशालिः दूषकः पुष्पाण्डकः पुण्डरीकः महिषमस्तकः दीर्घशूकः काञ्चनकः हायनः लोध्रपुष्पकः च तथा इत्याद्याः बहवः बहुदेशजाः सन्ति।
10. ग्रन्थविस्तरभीतेः ते समस्ताः अत्र न भाषिताः (सन्ति)।
11. शालयः मधुराः स्निग्धाः बल्याः बद्धाल्पवर्चसः कषायाः लघवः रुच्याः स्वर्याः वृष्याः बृंहणाः अल्पानिलकफाः शीताः पित्तघ्नाः तथा मूत्रलाः च (सन्ति)।
12. दग्धमृज्जाताः शालयः कषायाः लघुपाकिनः सृष्टमूत्रपुरीषाः रूक्षाः श्लेष्मापकर्षणाः च (सन्ति)।
13. कैदाराः (शालयः) वातपित्तघ्नाः गुरवः कफशुक्रलाः कषायाः अल्पवर्चस्काः मेध्याः बलावहाः च एव (सन्ति)।
14. स्थलजाः (शालयः) स्वादवः पित्तकफघ्नाः वातवह्निदाः किञ्चिद् तिक्ताः कषायाः विपाके कटुकाः च

## B Query Templates

| Category | Sanskrit Template | English Template | Input Type |
|---|---|---|---|
| सूचि (Contents) | पदार्थानां के प्रकाराः। | What are all the entity types? | |
| सूचि (Contents) | पदार्थेषु के सम्बन्धाः। | What are all the relationships? | |
| सूचि (Contents) | के के द्रव्याः। | What are all the substances? | |
| सूचि (Contents) | के के पदार्थाः। | What are all the entities? | |
| वर्णन (Detail) | {0} इत्यस्य विषये दर्शय। | Show some details about {0}. | Entity |
| वर्णन (Detail) | {0} इत्यस्य विषये अधिकं दर्शय। | Show some more details about {0}. | Entity |
| प्रकार (Type) | {0} इत्यस्य प्रकारः कः। | What is the type of {0}? | Entity |
| प्रकार (Type) | सर्वे {0} इति प्रकारस्य पदार्थाः चिनु। | Find all the entities of type {0}. | Entity-Type |
| गुण (Property) | केषां द्रव्याणां {0} इति गुणः अस्ति। | Which substances have a property {0}? | Entity |
| द्रव्य (Substance) | {0} इत्यस्य गुणाः के। | What are the properties of {0}? | Entity |
| द्रव्य (Substance) | {0} इति द्रव्यस्य प्रकाराः के। | What are the types/variants of the substance {0}? | Entity |
| समानार्थक (Synonym) | {0} इत्यस्य अन्यानि नामानि कानि। | What are the synonyms of {0}? | Entity |
| सम्बन्ध (Relation) | {0} इति सम्बन्धेन बद्धं सर्वं दर्शय। | Find all entities related by the relation {0}. | Relation |
| सम्बन्ध (Relation) | {0} {1} एतयोः मध्ये कः सम्बन्धः। | What is the relation between {0} and {1}? | Entity, Entity |

| Category | Sanskrit Template | English Template | Input Type |
|---|---|---|---|
| त्रिदोष (Tridoṣa) | के पदार्थाः {0} इति दोषस्य वर्धनं कुर्वन्ति। | Which entities increase the dosha {0}? | Entity |
| त्रिदोष (Tridoṣa) | के पदार्थाः {0} इति दोषस्य ह्रासं कुर्वन्ति। | Which entities decrease the dosha {0}? | Entity |
| त्रिदोष (Tridoṣa) | के पदार्थाः {0} इति दोषस्य वर्धनं {1} इति दोषस्य ह्रासं च कुर्वन्ति। | Which entities increase the dosha {0} and decrease the dosha {1}? | Entity, Entity |
| त्रिदोष (Tridoṣa) | के पदार्थाः {0} {1} एतयोः दोषयोः वर्धनं {2} इति दोषस्य ह्रासं च कुर्वन्ति। | Which entities increase the doshas {0} and {1} and decrease the dosha {2}? | Entity, Entity, Entity |
| त्रिदोष (Tridoṣa) | के पदार्थाः {0} इति दोषस्य वर्धनं {1} {2} एतयोः दोषयोः ह्रासं च कुर्वन्ति। | Which entities increase the dosha {0} and decrease the doshas {1} and {2}? | Entity, Entity, Entity |
| रोग (Disease) | के पदार्थाः {0} इति रोगं कुर्वन्ति। | Which entity causes the disease {0}? | Entity |
| रोग (Disease) | के पदार्थाः {0} इति रोगं हरन्ति। | Which entity cures the disease {0}? | Entity |
| प्रभाव (Effect) | के पदार्थाः {0} एतं विकुर्वन्ति। | Which entities affect {0}? | Entity |
| प्रभाव (Effect) | के पदार्थाः {0} एतस्मै लाभप्रदाः। | Which entities benefit {0}? | Entity |
| प्रभाव (Effect) | के पदार्थाः {0} एतस्मै क्षतिप्रदाः। | Which entities harm {0}? | Entity |
| प्रभाव (Effect) | के पदार्थाः {0} इत्यस्य वर्धनं कुर्वन्ति। | Which entities increase {0}? | Entity |
| प्रभाव (Effect) | के पदार्थाः {0} इत्यस्य ह्रासं कुर्वन्ति। | Which entities decrease {0}? | Entity |
| अधिकरण (Space-Time) | {0} इति पदार्थः कदा जायते। | When does {0} grow? | Entity |
| अधिकरण (Space-Time) | {0} इति पदार्थः कुत्र लभ्यते। | Where is {0} found? | Entity |
| साधारण (Generic) | के पदार्थाः {0} इति पदार्थेन सह {1} इति सम्बन्धेन सम्बन्धिताः। | Which entity is related to {0} by relation {1}? | Entity, Relation |
| साधारण (Generic) | {0} इति पदार्थः {1} इति पदार्थेन सह कथं सम्बन्धितः। | How is {0} related to {1}? | Entity, Relation |
| साधारण (Generic) | {0} इति प्रकारस्य पदार्थैः सह {1} इति सम्बन्धेन बद्धाः {2} इति प्रकारस्य पदार्थान् दर्शय। | Show all matches where an entity of type {0} has relation {1} with an entity of type {2}. | Entity-Type, Relation, Entity-Type |